# A Random Walk Based Algorithm for Structural Test Case Generation

Jifeng Xuan[1], He Jiang[2], Zhilei Ren[1], Yan Hu[2], Zhongxuan Luo[1, 2]

[1] School of Mathematical Sciences
Dalian University of Technology
Dalian, China
{xuan, ren}@mail.dlut.edu.cn

[2] School of Software
Dalian University of Technology
Dalian, China
{jianghe, huyan, zxluo}@dlut.edu.cn

*Abstract*—Structural testing is a significant and expensive process in software development. By converting test data generation into an optimization problem, search-based software testing is one of the key technologies of automated test case generation. Motivated by the success of random walk in solving the satisfiability problem (SAT), we proposed a random walk based algorithm (WalkTest) to solve structural test case generation problem. WalkTest provides a framework, which iteratively calls random walk operator to search the optimal solutions. In order to improve search efficiency, we sorted the test goals with the costs of solutions completely instead of traditional dependence analysis from control flow graph. Experimental results on the condition-decision coverage demonstrated that WalkTest achieves better performance than existing algorithms (random test and tabu search) in terms of running time and coverage rate.

## I. INTRODUCTION

Software testing is an expensive and time-consuming process occupying about 50% resource of software development cycle [1][25]. A main task of software testing is to generate test cases to detect errors. As it is impossible to execute a program exhaustively, the test coverage and adequacy criterion are employed to measure the quality of software testing by a fraction of test cases [2]. Structural testing (white-box testing) generates test cases from the codes of a program. According to different goals, the criterion of structural testing provides distinct granularity, including statement coverage, branch coverage, multiple-condition coverage, and path coverage [2]. Among these coverage criteria, statement coverage is easy to achieve since it only covers every statement of a program; and branch coverage covers every branch to ensure the correctness of basic structures in a program. Some other criteria (e.g., path coverage) provide stronger guarantee than branch coverage, but far more test cases are necessary.

Search-based software testing is a dynamic approach for automatic test case generation [3]. In contrast to traditional static approach (symbolic execution), search-based software testing can efficiently reduce the running time and space requirement. The kernel of this search-based technology is to convert the test data generation into an optimization problem and to solve it with approximate optimization algorithms. Miller & Spooner [4] firstly treated test case generation as an optimization problem and used a hill-climbing algorithm on float-point data in 1976; Korel [5] extended their work and established a prototype of modern dynamic approaches in 1990. Aside from these traditional optimization algorithms, numerous meta-heuristic algorithms have received widespread research interests in test case generation, including genetic algorithm by Xanthakis, et al. [6] and Michael, et al. [7], simulated annealing by Tracey, et al [8], scatter search by Sagarna & Lozano [9], particle swarm optimization by Windisch, et al. [11], estimation of distribution algorithms by Sagarna, et al. [21], and tabu search by Díaz, et al. [10]. All these optimization algorithms are applied to generate test cases collaborated with a set of objective functions. An objective function is designed to adapt to a test goal, which is usually defined as an element in codes, such as a branch in branch coverage or a path in path coverage. Some of objective functions are effective in the application of test case generation, but are complex and hard to understand or implement [10][16][20]. McMinn surveys the objective functions of search-based testing in [3].

Search-based software testing is an approximate approach, which usually searches test cases incorporated some randomized mechanism. In random process, random walk is a trajectory that consists of successive random steps [22]. Random walk has been applied in many fields of computer science, including information retrieval [23], machine learning [24], and constraint programming [13][14][15]. In constraint programming, local search algorithms based on random walk exhibits strong ability in approximate solving satisfiability problem (SAT) and its variants [13][14]. Motivated by this fact, we considered test case generation as random walk of test cases and proposed a random walk based algorithm (called WalkTest) to solve the automatic structural test case generation problem.

WalkTest iteratively calls a random walk operator to obtain high quality solutions in a solution pool. Firstly, WalkTest encodes solutions with Gray codes instead of natural binary codes and converts the test goals into minimum objective functions. Then, WalkTest selects and updates solutions based on a probabilistic walk on the continuous search space of Gray codes. After that, the test goals (objective functions) are sorted by minimum cost rather than the dependence analysis from control flow graph (CFG) to reduce the running time. Finally,

This work is partially supported by the Natural Science Foundation of China under Grant No. 60805024, the National Research Foundation for the Doctoral Program of Higher Education of China under Grant No. 20070141020.

WalkTest collects the statistics of coverage and test cases. Experimental results on typical programs indicated that WalkTest achieves a high coverage rate of structural testing (condition-decision coverage) in less time than existing algorithms (random test and tabu search). These results also demonstrated that WalkTest was not sensitive to the values of input parameters, especially in terms of running time.

The remainder of this paper is organized as follows. Section II introduces structural test case generation and related works. Section III describes WalkTest algorithm in detail, including the framework, its local search operator, and the sorting strategy of test goals. Experimental results are presented in Section IV. Finally, Section V concludes this paper and discusses the directions of future work.

## II. SEARCH-BASED STRUCTURAL TESTING

Structural testing focuses on the implementation details of program units. Among the coverage criteria of structural testing, branch coverage (also known as decision coverage) is discussed by most literatures [2][5][6][9][11]. This criterion meets a requirement that every branch with conditions must be executed at least once. A branch is one possible path in the execution for a branch node, which is defined as some continuous codes of a selection structure. Usually, one branch consists of some conditions with connection of logical operators. Fig. 1 illustrates an example of a fraction of codes with two branches for an *if* statement. For the branches with more than one condition, branch coverage cannot predicate the satisfiability of every condition. Therefore, we choose condition-decision coverage (C/D coverage) [7][10] as the criterion to achieve the test cases for each conditions in this paper.

Besides the requirement of branch coverage, C/D coverage requests the execution of both TRUE and FALSE values in every condition, i.e., every condition must be evaluated TRUE or FALSE at least once for C/D coverage. Table I gives the requirements of test case generation in C/D coverage for the instance in Fig. 1. In addition to Line 1 and 2 for branch coverage, C/D coverage is satisfied, if and only if all eight lines are satisfied by a set of test cases. Moreover, if a test case has been designed to satisfy Line 1 in Table I, Line 3, Line 5, and Line 7 can be done simultaneously due to the logical inclusion relationship of Line 1. Thus, this relationship between a branch and its conditions leads to a reduction of test goals without loss of test quality.

Under a coverage criterion, the test goals are transformed into the value of objective functions and the optimal solutions for these functions are transformed back into test cases in search-based technology. For objective functions in test data generation, a uniform approach is to design a minimum function with nonnegative values [3][8]. In other words, the conditions or conditional branches in codes are converted into functions to indicate the coverage with the values at least zero. The forms of objective functions depend on the search strategies in different algorithms. We choose a prior and simple set of objective functions described in [8] to perform WalkTest. Table II presents the details of these functions in C++ language. Every operator (logic operator, e.g., && or relation operator, e.g., < ) in test goals is uniformly defined as a piecewise function. The value of this function equals to zero while the logical expression is satisfied by a test case; otherwise, it equals to a positive value to predicate the distance between the current and optimal test cases. For example, a condition "a < b" with operator "<" is defined as a function, which equals to zero, if and only if a < b, or equals to a value of (b - a) +K. An exception is the operator "*logical negation*" (i.e., "*!*" in C++ language), which is moved inwards and propagated over the original expression rather than defining an extra function. For unsatisfied logical expressions, the function values illustrate the distance to the optimal solutions.

## III. RANDOM WALK SEARCH

In this section, we present WalkTest for automated structural test case generation in details: framework, random walk operator, and sorting strategy.

Before the technical details of WalkTest, we give a brief study on a classic program "triangle classifier". This program contains only 6 branch nodes, and well-known as a benchmark for structural testing [19]. Fig. 2 shows the control flow graph of "triangle classifier". Close dependence exists among these nodes while a test case is generated to cover some of them. For example, Nodes 0, 2, 4, 5 must be covered as a precondition for the condition "B==C" in Node 6. This dependence leads to two difficulties. On one hand, there is no any direct relationship between the conditions of Node 4 and the condition of Node 6. Thus, during the process of search, Node 4 cannot provide a direct guide for Node 6. On the other hand, Node 6, the last node of dependence, cannot be easily covered by random test (if the input variables are 32-bit integers, the coverage probability of Node 6 is $1/2 \times 2^{32}/(2^{32})^3$, i.e., $2^{65}$). For these reasons, it is necessary to find a search algorithm to handle such difficult nodes in test case generation.

```
1    if ( A > 0 && B > 0 && C > 0 )
2        target 1;
3    else
4        target 2;
```

Figure 1. An example of codes in a branch node. The first two lines and the last two lines are both branches. In the first branch, there are three conditions connected by *logical and "&&"* operators (*C++* language).

TABLE I. REQUIREMENTS OF TEST CASE GENERATION IN C/D COVERAGE

| Line | Requirements of test case generation |
|------|--------------------------------------|
| 1    | A > 0 && B > 0 && C > 0              |
| 2    | ! ( A > 0 && B > 0 && C > 0 )        |
| 3    | A > 0                                |
| 4    | ! ( A > 0 )                          |
| 5    | B > 0                                |
| 6    | ! ( B > 0 )                          |
| 7    | C > 0                                |
| 8    | ! ( C > 0 )                          |

TABLE II. OBJECTIVE FUNCTIONS IN C++ LANGUAGE

| Expression | Objective function ( *fit* () ) |
|---|---|
| Boolean | if *TRUE* then 0 else $K$ |
| a == b | if a == b then 0 else $abs$(a - b) + $K$ |
| a != b | if a != b then 0 else $K$ |
| a < b | if a < b then 0 else (a - b) + $K$ |
| a <= b | if a <= b then 0 else (a - b) + $K$ |
| a && b | min ( *fit* (a), *fit* (b) ) |
| a \|\| b | *fit* (a) + *fit* (b) |
| ! a | Negation is moved inwards and propagated over a |

$K$ is defined as a minimal positive number. Function *abs*( a ) is the absolute value of a.

| Algorithm 1: Framework of WalkTest |
|---|
| **Input:** variable list $L$, times $r$ and $t$, solution pool $P$ with size $q$, set of test goals $T$ |
| **Output:** coverage rate $R$, test case set $X$ |
| 1　**while** $R$ < 100% and the iteration times < $r$ **do** |
| 2　　Initialize the solution pool $P$ with $q$ and $i = 0$ ; |
| 3　　Run random test for $L$ |
| 4　　　and update $P$ for $t$ times with best solutions; |
| 5　　**while** $i$ < number of test goals **do** |
| 6　　　Sort all test goals and select one test goal $T_i$ ; |
| 7　　　Search solutions for $T_i$ by random walk and update $P$; |
| 8　　　$i = i + 1$; |
| 9　　**endwhile** |
| 10　**endwhile** |
| 11　Calculate coverage rate $R$ and test case set $X$; |

some initial solutions. Based on these solutions, all unsolved test goals are sorted to specify the solving priority. More details of this sorting strategy will be discussed in Part C of Section III. After the sorting, the first test goal is selected for a random walk operator (see Part B of Section III). This operator detects the local optimal solution over a series of trials on the binary form of variables. Finally, a final coverage rate is calculated and the test case set for this coverage is collected.

A solution pool is incorporated into WalkTest to record good quality solutions in search history. This pool stores the solutions with minimum costs for every test goal. Due to the capacity limitation of solution pool, the anterior solutions with bad quality can be flushed out by the posterior good solutions. In addition to recording good solutions, this solution pool is employed to update solutions in it. Some literatures [9][17] have reported that handling one test goal can guide the process for other goals. Thus, a test goal may be improved or solved as a "side effect", when the search algorithm tries to handle another goal. By this strategy, all the solutions in the pool are updated while the related test goals are not achieved in the search algorithm. Before such search algorithm, the random test mentioned above is assigned to fill some solutions into the pool. In general, random test can cover some of test goals without search algorithms.

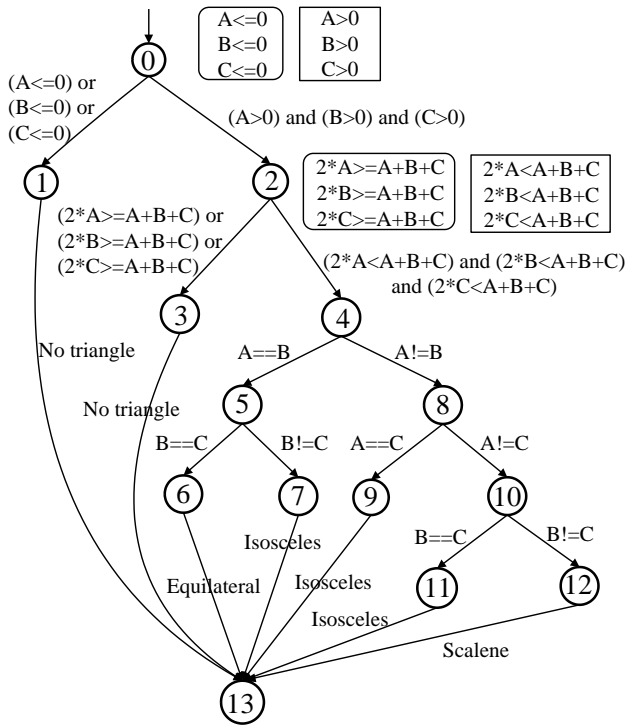

Figure 2. Control flow graph of the "triangle classifier" program. This program distinguishes the type of triangle according to the three input edges (e.g., Node 6 indicates an equilateral triangle).

In this paper, we employ random walk to achieve the randomness in test case generation. Our algorithm, WalkTest, searches the solutions similar as walking in the search space randomly. The walk steps are controlled by a probability parameter. The framework of WalkTest iteratively calls a random walk operator to search solutions and sorts test goals for the next search iteration.

*A. Framework of WalkTest*

For an input program, the test goals are converted into a set of objective functions under a coverage criterion. After this conversion, WalkTest generates test cases for the input variables with these objective functions. We describe the framework of WalkTest in Algorithm 1. WalkTest consists of a series of iterations as a multi-restart strategy. At the beginning of every iteration, a simple random test is employed to produce

*B. Random Walk Operator*

We employ a random walk search algorithm as a local search operator in WalkTest. This random walk operator treats the process of searching for test cases as a random walk in search space. The search process starts by choosing a solution in the solution pool. Then this operator selects a binary bit of this solution, and updates it from TRUE to FALSE or vice versa. This action is repeated until no better solutions can be explored. To escape from the monotonous solutions assembled near the original solutions, a probability is specified to update the solutions from some ones with bad quality.

Given a program in structural testing, the solution is formed by a vector of values for input variables. In our algorithm, this vector is encoded to a binary string, which is a connection of binary form of all separate variables. Thus, only one binary string needs to be constructed for a program. To provide a continuous search space, we encode the solutions with Gray codes. Gray code has been applied in some optimization algorithms to overcome the hardness of continuous search [3][18]. Gray code is formed as a code with 1-Hamming distance (Hamming distance is defined as the number of different bits between two solutions) in Hamming space. In random walk, a flip is defined as one bit from 1 to 0 or vice versa in changing one binary string. Table III shows the difference between natural binary code and Gray code.

Algorithm 2 presents the details of random walk operator in WalkTest. The random walk operator iteratively flips Gray codes of variable vector to generate test cases for specific test goals. This operator randomly selects a solution from the solution pool and flips it repeatedly. When the pool is empty, a random solution is generated as the initial solution instead of the above selection. After flipping every bit of the selected solution, a set of candidate solutions is formed. The random walk operator selects a subset of solutions with the minimum cost. Then the operator flips the solutions under the following rules: if the cost is better than the original one, the operator updates the solution and cost; otherwise, it updates the solution from the universal set or the subset according to a specific probability. During all the steps, the solution pool is always updated to record good solutions.

This random walk operator contains five random actions (in Line 3, 4, 14, 15, and 16 of Algorithm 2, respectively), which are designed to provide the diversity of solutions. The last two of these actions are controlled under a specific probability. The approach of evaluating the value of this probability is discussed in [12][13]. Although a dynamic strategy for the probability leads to a rapid convergence of algorithm, it introduces some additional parameters and complex implementation. Thus, we simply set it to a static value in our algorithm.

## C. Test Goals Sorting

Before the application of random walk operator, the unsolved test goals are sorted in the search framework. Some literatures discussed the strategies about this sorting [9][10][17]. In these literatures, the sorting approaches focus on the CFG. For instance, Díaz, et al. [10] provide a strategy that a node is picked as a test goal after its parent node is covered in CFG; Sagarna & Lozano [9] distinguish the priority of test goals with the same cost by the breadth-first traversal in CFG.

Since the traversal of a graph (such as CFG) is expensive both in implementation and running time, we attempt to sort goals only related to the costs of objective functions. The weight of a goal is defined as the minimum cost among solutions of this test goal in solution pool. Meanwhile, the test goals without any solution in the pool are set to a maximum cost. WalkTest sort the test goals with the weights.

TABLE III. AN 8-BIT FORM OF NATURAL BINARY CODE AND GRAY CODE

| Decimal value | Natural binary code | Gray code |
|---|---|---|
| 7 | 00000111 | 00001000 |
| 8 | 00000100 | 00001100 |

Decimal value 7 is only 1 bit away from 8 in Gray code, but 4 bits in natural binary code. Thus, at least 4 flips are needed to switch between decimal 7 and 8 in natural binary code, which may be only 1 flip in Gray code.

A test goal with the least weight is easier to solve than any other unsolved test goal. This can be explained as follows. For two test goals A and B, if handling A can lead B to be solved, it is certain that A has a less weight than B. B cannot be covered unless A is covered within one test case. Furthermore, if A and B can not provide a guidance for each other, the weight of them represents the distance between the current solution and the optimal solution. Based on this property, the weight of test goals will give a hint for the hardness of resolving. In addition, the goals with the same weight within one sorting will be sorted by the number of solutions in the pool. Evidently, the goals with more solutions are more likely to converge to the optimal solution in the search process. It is unnecessary to provide the dependence of test nodes in a graph form.

## IV. EXPERIMENTAL RESULTS

In this section, we demonstrate the experimental results over some classic programs. WalkTest is implemented in $C++$ and compiled in $g++$ under an *Intel Pentium D 2.8* GHz with *1 GB* memory running *Fedora 9* (*Linux Kernel 2.6*). The experiments also run under this environment.

In comparison with the best heuristic (tabu search [10]) for C/D coverage, we conduct our experiments on the same widely used programs, including the "triangle classifier program with integer", the "triangle classifier program with real number", the "line rectangle classifier" and the "number of days between two dates" with the same implement as [10]. The "triangle classifier" problems are to recognize the type of triangle by the three input edges (its CFG is in Fig. 2); the "line rectangle classifier" problem is to obtain the relationship of position between an input line and an input rectangle; and the "number of days between two dates" problem is to calculate the number of days between the two input dates. The key characteristics of every test problem are summarized in Table IV.

---

**Algorithm 2**: Random walk operator

**Input:** test goal $T_i$, times $m_1, m_2$, probability $p$, solution pool $P$
**Output:** the updated pool $P^*$

1   **while** times of running $< m_1$ **do**
2     **if** a solution of $T_i$ in $P$ exists
3     **then** select solution $s_i$ randomly in $P$;
4     **else** generate one solution $s_i$ randomly;
5     **endif**
6     Record the cost $c_i$ of $s_i$;
7     Generate Gray code $g_i$ of $s_i$;
8     **while** walking less than $m_2$ times **do**
9       Flip every bit of $g_i$ to generate a set of solutions $S_i$;
10      Calculate the cost of solutions in $S_i$;
11      Update $P$ to $P^*$ with new solutions with better quality;
12      Select a subset $S_i^*$ of optima in $S_i$, with cost $cs_i$;
13      **if** $cs_i < c_i$
14      **then** pick one solution $g_i^*$ in $S_i^*$, randomly;
15      **else** pick a solution $g_i^*$, with probability $p$ from $S_i$,
16        or with $1-p$ from $S_i^*$;
17      **endif**
18      Update $g_i$ with $g_i^*$ and update $c_i$ with the cost of $g_i^*$;
19     **endwhile**
20 **endwhile**

TABLE IV. THE CHARACTERISTICS OF TEST PROGRAMS

| Program | Node | B-node | Obj. | Loop | Var. | Type |
|---|---|---|---|---|---|---|
| Tri-int | 12 | 6 | 24 | 0 | 3 | Int |
| Tri-real | 12 | 6 | 24 | 0 | 3 | Real |
| Line-rect | 53 | 18 | 98 | 0 | 8 | Real |
| Day-date | 123 | 42 | 108 | 3 | 6 | Int |

The column "Node" and "B-node" show the number of nodes and branch nodes of every program. The following three columns give the number of test goals in C/D coverage, loops, and input variables. And the last column is the type of input variables with "Int" as integers and "Real" as real numbers.

In the experiments of WalkTest, we set both the parameters r and t to 100 in Algorithm 1, respectively, and set both m1 and m2 to 5 in Algorithm 2. The size of solution pool $q$ is 40 and the probability parameter $p$ is 2/3. WalkTest repeatedly runs for 100 times to achieve the average time and the maximum time of running. TSGen and random test are used for comparison. All the results and parameter values of TSGen are collected from [10]. In random test, we follow [10] to generate 10,000,000 test cases randomly.

Fig. 3 indicates the relationship between running time and coverage rate of four programs in Table IV. In the experiment, we compare the running time and coverage rate of the three algorithms mentioned above. WalkTest and random test keep steady tendency in this figure. For coverage rate, WalkTest always stays above 90%, but the random test stays on a low level. TSGen cannot achieve high coverage rate in short time as WalkTest, although TSGen keeps ascending in the curve. To achieve the same coverage rate, WalkTest consumes less time than the other two reported algorithms, especially random test.

All the experimental results are summarized in Table V. WalkTest achieves a notably improvement over the other two approaches on C/D coverage. The coverage rates of all test programs are easily achieved 100% by WalkTest. Besides the average running time, the maximum running time of WalkTest is less than the average time of the other two algorithms. The running time of WalkTest is far less than that of random test, i.e., under 30% of running time of random test. WalkTest also cost less time than TSGen, i.e., from 3% to 98% of running time of TSGen. In WalkTest, there is a smooth increase from the average running time to the maximum running time for most of programs. However, a sharp increase exists in the experiments for the "line rectangle classifier" problem in running time. It implies that the running time of WalkTest is instable for some programs.

For the programs of the same input variables with different precisions, WalkTest consumes nearly the same running time. For the programs with different variable types, WalkTest also exhibits the similar phenomenon. It implies that WalkTest is not sensitive to the data type or precision of input variables.

V. CONCLUSION

In this paper, we proposed a random walk based algorithm (WalkTest) to solve the problem of structural test case generation. After converting the test case generation into an optimization problem, WalkTest provides a framework with a sorting strategy of test goals and a random walk operator. The sorting strategy presents a priority for test goals, which is implemented by costs of goals instead of traditional dependence analysis from control flow graph. The random walk operator works iteratively to achieve local optimal solutions. In all the programs of our experiments, 100% coverage rate is easy to achieve in a short time. Although WalkTest is designed and tested on condition-decision coverage, it can be applied on some other coverage criteria, such as branch coverage or path coverage.

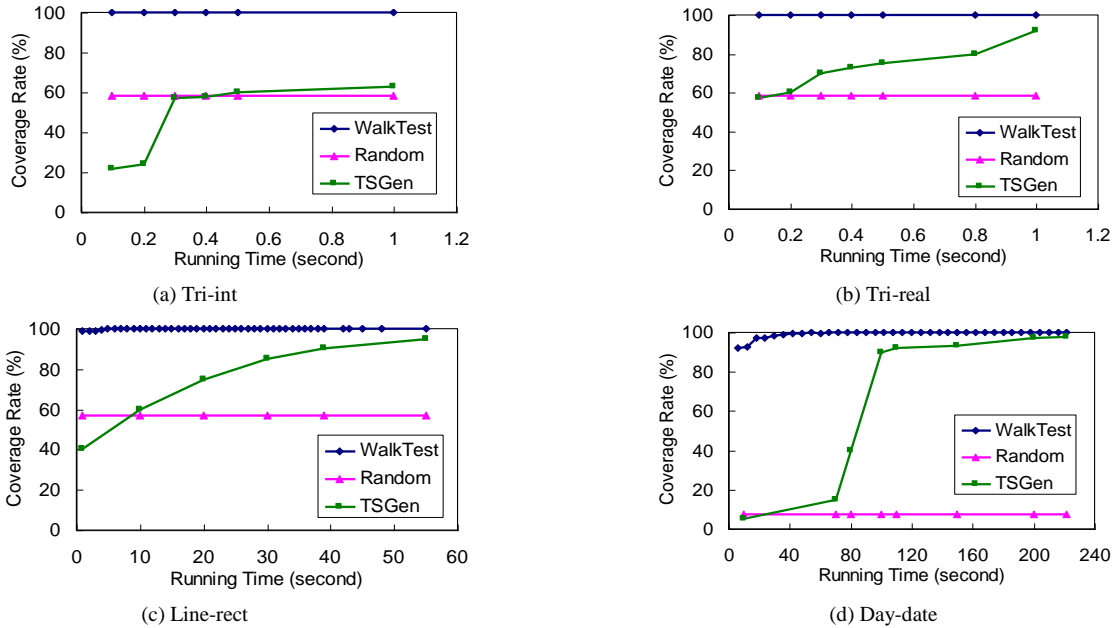

(a) Tri-int  (b) Tri-real  (c) Line-rect  (d) Day-date

Figure 3. Relationship between running time and coverage rate. The x-axis is the running time in seconds and the y-axis is the coverage rate in percentage.

TABLE V. EXPERIMENTAL RESULTS FOR PROGRAMS

| Program | Rand | | TSGen | | WalkTest | | | | |
|---|---|---|---|---|---|---|---|---|---|
| | %cov | Ave. | %cov | Ave. | %cov | Ave. | Max. | %Rand | %TSGen |
| Tri-int, 32 bits | 58.33 | 298 | 100 | 21.43 | 100 | 0.80 | 2.00 | 99.73 | 96.27 |
| Tri-real, ±100,000.000 | 58.33 | 330 | 100 | 0.86 | 100 | 0.84 | 1.96 | 99.75 | 2.33 |
| Tri-real, ±2,000,000.0000 | 58.33 | 340 | - | - | 100 | 0.96 | 2.21 | 99.72 | - |
| Line-rect, ±100,000.000 | 57.14 | 1210 | 100 | 60.69 | 100 | 17.52 | 53.24 | 98.55 | 71.13 |
| Line-rect, ±2,000,000.0000 | 57.14 | 1250 | - | - | 100 | 20.72 | 68.64 | 98.34 | - |
| Day-date, 32 bits | 7.41 | 686 | 100 | 251.38 | 100 | 204.84 | 221.28 | 70.14 | 18.51 |

The column "Program" indicates the four programs in Table IV and the range after the program's name shows the digital precision, e.g., "32 bits" means every input variable is an integer with 32 bits; "±100,000.000" means a real number with the range ±100,000 and three significant digits after the point. The following three columns show the performance and results of the three algorithms, respectively. The sub-columns: namely "%cov" shows the coverage rate in percentage and "Ave." shows the average running time in seconds. Column "WalkTest" has three extra sub-columns as follows: "Max." shows the maximum running time in seconds, "%Rand" and "%TSGen" show the reduction of average running time in percentage over random test and TSGen (e.g., the value of %Rand is defined as 1 − Ave. of WalkTest / Ave. of Rand ).

In future work, some modifications can be added to WalkTest to enhance its performance. For an instance, we will combine the static probability parameter with a self-adaptive strategy driven by the values of objective functions. In addition, we will apply WalkTest to the object-oriented software testing in our future work.

ACKNOWLEDGMENT

We thank Dr. Eugenia Díaz for sharing the source code of the three programs. Dr. Eugenia Díaz is with Department of Computer Science, University of Oviedo in Spain.

REFERENCES

[1] B. Beizer, Software Testing Techniques, 2nd ed. New York, NY: Van Nostrand Reinhold, 1984.
[2] H. Zhu, P. A. V. Hall, and J. H. R. May, "Software unit test coverage and adequacy," ACM Computing Surveys, vol. 29, no. 4, pp. 366-427, Dec. 1997.
[3] P. McMinn, "Search-based software test data generation: a survey," Software Testing, Verification and Reliability, vol. 14, no. 2, pp. 105-156, Jun. 2004.
[4] W. Miller and D. L. Spooner, "Automatic generation of floating-point test data," IEEE Trans. Software Engineering, vol. 2, no. 3, pp. 223-226, May 1976.
[5] B. Korel, "Automated software test data generation," IEEE Trans. Software Engineering, vol. 16, no. 8, pp. 870-879, Aug. 1990.
[6] S. E. Xanthakis, C. C. Skourlas, and A.K. LeGall, "Application of genetic algorithms to software testing," in Proc. 5th Int. Conf. Software Engineering and its Applications, Toulouse, France, 1992, pp. 625-636.
[7] C. C. Michael, G. McGraw, and M. A. Schatz, "Generating software test data by evolution," IEEE Trans. Software Engineering, vol. 27, no. 12, pp. 1085-1110, Dec. 2001.
[8] N. Tracey, J. A. Clark, K. Mander, and J. A. McDermid, "An automated framework for structural test-data generation," in Proc. 13th IEEE Conf. Automated Software Engineering (ASE 98), Honolulu, HI, 1998, pp. 285-288.
[9] R. Sagarna, and J. A. Lozano, "Scatter search in software testing, comparison and collaboration with estimation of distribution algorithms," European Journal of Operational Research, vol. 169, no. 2, pp. 392-412, Mar. 2006.
[10] E. Díaz, J. Tuya, R. Blanco, and J. J. Dolado, "A tabu search algorithm for structural software testing," Computers and Operations Research, vol. 35, no. 10, pp. 3052-3072, Oct. 2008.
[11] A. Windisch, S. Wappler, and J. Wegener, "Applying particle swarm optimization to software testing," in Proc. 9th Conf. Genetic and Evolutionary Computation (GECCO 07), London, England, 2007, pp. 1121-1128.
[12] H. H. Hoos, "An adaptive noise mechanism for WalkSAT," in Proc. 18th National Conference on Artificial Intelligence (AAAI 02), Edmonton, Canada, 2002, pp. 655-660.
[13] B. Selman, H. A. Kautz, and B. Cohen, "Noise strategies for improving local search," in Proc. 12th National Conference on Artificial Intelligence (AAAI 02), Seattle, WA, 1994, pp. 337-343.
[14] W. Zhang, A. Rangan, and M. Looks, "Backbone guided local search for maximum satisfiability," in Proc. 18th Int. Joint Conf. Artificial Intelligence (IJCAI 03), Acapulco, Mexico, 2003, pp. 1179-1186.
[15] H. Jiang, J. Xuan, and Y. Hu, "Approximating backbone in the weighted maximum satisfiability problem," in Proc. 1st Int. Symp. Parallel Algorithms, Architectures and Programming, Hefei, China, 2008, pp. 80-93.
[16] J. Wegener, A. Baresel, and H. Sthamer, "Evolutionary test environment for automatic structural testing," Information and Software Technology, vol. 43, no. 14, pp. 841-854, Dec. 2001.
[17] R. Sagarna and X. Yao, "Handling constraints for search based software test data generation," in Proc. IEEE Int. Conf. Software Testing Verification and Validation Workshop (ICSTW 08), Lillehammer, Norway, 2008, pp. 232-240.
[18] D. Whitely, S. Rana, J. Dzubera, and K. E. Mathias, "Evaluating evolutionary algorithms," Artificial Intelligence, vol. 85, no. 1-2, pp. 245-276, Aug. 1996.
[19] G. J. Myers, C. Sandler, T. Badgett, and T. M. Thomas, The Art of Software Testing, 2nd ed.  New York, NY: John Wiley & Sons Inc., 2004, pp. 1-2.
[20] L. Bottaci, "Predicate expression cost functions to guide evolutionary search for test data," in Proc. 5th Conf. Genetic and Evolutionary Computation (GECCO 03), Chicago, IL, 2003, pp. 2455-2464.
[21] R. Sagarna, A. Arcuri, and X. Yao, "Estimation of distribution algorithms for testing object oriented software," in Proc. IEEE Congress Evolutionary Computation (CEC 07), Singapore, 2007, pp. 438-444.
[22] P. G. Doyle and J. L. Snell, Random Walks and Electric Networks. Washington, DC: Mathematical Association of America, 1984, pp. 3-5.
[23] Z. Bar-Yossef and M. Gurevich, "Random sampling from a search engine's index," in Proc. 15th Int. Conf. World Wide Web (WWW 06), Edinburgh, Scotland, 2006, pp. 367-376.
[24] K. Toutanova, C. D. Manning, and A. Y. Ng, "Learning random walk models for inducing word dependency distributions," in Proc. 21st Int. Conf. Machine Learning (ICML 04), Banff, Canada, 2004, pp. 13.
[25] Y. Peng, G. Kou, G. Wang, H. Wang, and F. Ko, "Empirical Evaluation Of Classifiers For Software Risk Management," Int. Jour. Information Technology and Decision Making (IJITDM), vol. 8, no. 4, pp. 749-768, 2009.